\documentclass[letterpaper, 12pt]{article}

\usepackage{booktabs}
\usepackage{listings}

\usepackage{setspace}
\usepackage{amsmath}
\usepackage{amssymb}

\usepackage{xcolor}

\usepackage{hyperref}
\usepackage{tikz}

\usepackage{amsthm}
\usepackage{lipsum}

\theoremstyle{remark}

\newtheorem{assumption}{Assumption}

\usepackage[top=1.2in, bottom=1.18in, left=1.28in, right=1.28in, letterpaper]{geometry}
\usepackage{JASA_manu}

\newcommand{\param}{\boldsymbol{\theta}}

\theoremstyle{definition}
\newtheorem{example}{Example}

\usepackage{natbib}
\title{Evaluating the quality of survey and administrative data with generalized multitrait-multimethod models\thanks{The authors are indebted to Hal Stern and J\"org Drechsler for their comments as well as Barbara Felderer for her assistance in preparing the data. This work was supported by the Netherlands Organization for Scientific Research (NWO) [Veni grant number 451-14-017].}}
\author{DL Oberski$^1$  \and A Kirchner$^{2}$  \and S Eckman$^{3}$  \and F Kreuter$^{4,5,6}$}

\date{
\begin{footnotesize}
$^1$ Tilburg University, The Netherlands\\
$^2$ University of Nebraska, United States \\
$^3$ RTI International, United States \\
$^4$ Institute for Employment Research, Germany \\
$^5$ University of Mannheim, Germany \\
$^6$ University of Maryland, United States\\
\end{footnotesize}
}
\begin{document}
	
\maketitle

\begin{abstract}
Administrative register data are  increasingly important in statistics, but, like other types of data, may contain measurement errors. To prevent such errors from invalidating analyses of scientific interest, it is therefore essential to estimate the extent of measurement errors in administrative data. Currently, however, most approaches to evaluate such errors  involve either prohibitively expensive audits or comparison with a survey  that is assumed perfect.

We introduce the ``generalized multitrait-multimethod'' (GMTMM) model, which can be seen as a general framework for evaluating the quality of administrative and survey data simultaneously. This framework allows both survey and register to contain random and systematic measurement errors. Moreover, it accommodates common features of administrative data such as discreteness, nonlinearity, and nonnormality, improving similar existing models. The use of the GMTMM model is demonstrated by application to linked survey-register data from the German Federal Employment Agency on income from and duration of employment, and a simulation study evaluates the estimates obtained.

\end{abstract}

\if 1=2
\section*{TODO}

\begin{itemize}
	\item Maybe emphasize more models that do not exist yet in the  examples.
	\item Apply different model to example? Censored and/or heteroskedastic for income, gamma for duration.
	\item Do we need a simulation?
	\item Sec 2 use multiple sources to estimate error $\rightarrow$ MTMM, but admin not normal $\rightarrow$ sec 3
	\item Include 3.3 in sec 2
	\item p4 Method by Person interaction, measurement effect differs over people
	\item sec 3 straight to equations
	\item Fig 1 connect to text esp eons; explain conventions used
	\item Sec leave out covariates
	\item[x] Application: Only employed residents $\backslash$ civil servants
	\item Note restrictions
	\item[x] Application: examples of errors are monthly data from yearly dbs, bonuses counted in year by admin, but year+1 by person
	\item[x] Give $n$ in appendix
\end{itemize}
\fi

\section{Introduction}\label{sec:intro}

Register data and administrative records play an increasingly important role in statistics \citep{wallgren2007register}, and several authors recommend and predict the increased use of ``big data''   \citep{entwistle2013new,podesta2014bigdata}, including administrative register data \citep{japec2015aapor}. Uses to date include studies of how agricultural households affect land changes \citep{rindfuss2004developing}, voter turnout \citep{ansolabehere2012validation}, or how peoples' numerical ability relates to mortgage default \citep{gerardi2013numerical}. However, there is evidence that register data may contain considerable measurement errors \citep{groen2012sources}. For example, \citet[p. 15]{bakker2012estimating} estimated that 24\% of the variance in Dutch official hourly wages records was random measurement error; \citet[p. 1]{ansolabehere2010quality} reported that 16.1 million out of the 185.4 million listed voter registration records in the United States were invalid; and \citet[p. 275]{ladouceur2007robustness} suggested that 20\% to 30\% of osteoarthritis cases are not registered in Quebec hospital administrative records, causing bias in prevalence estimates. The measurement error present in administrative records can severely bias and invalidate research results \citep{carroll2006measurement,saris_design_2007,vermunt2010latent}. It is therefore essential to evaluate the extent of measurement error in register data.\footnote{We  use the terms ``register data'' and ``administrative data'' synonomously to avoid repetition.}

The difficulty in studying error in register and administative data, however, is that there is often no ``gold standard'' measure. Some authors have suggested to link administrative registers to a survey, assuming the survey contains no measurement error \citep[e.g.][]{yucel2005imputation}. But measurement error in survey data is widespread \citep{hansen1961measurement,hansen1964estimation,felligi1964response,andrews1984construct,alwin_margins_2007,saris_design_2007,biemer2011latent}, and is in fact often measured by taking administrative records as the ``gold standard'' \citep[e.g.][]{kapteyn2007measurement,kreuter2010nonresponse,sakshaug2010nonresponse,kim2014response}. Thus, we often have two data sources, both measured with error, and we are interested in estimating the error in both.

Very few studies have attempted to estimate measurement error in both survey and administrative data simultaneously. \citet{nordberg2004measurement} discussed a longitudinal latent Markov model of measurement error in income, but again assumed the administrative register to be perfect in cross-sectional data; \citet{pavlopoulos2013measuring} applied a similar latent Markov model to unemployment data;  and \citet{bakker2012estimating} and \citet{scholtus2015npso} estimated measurement error using linear factor analysis. However, the models used in these studies have several drawbacks when applied to administrative register data.  First, true values of the variables of interest are often censored, zero-inflated, gamma, count, or nominal, and thus models which assume normally distributed true values are not appropriate. For example, income is usually zero-inflated and occupation is nominal. Second, the measurement error process in registers is likely to lead to nonnormal and nonlinear errors, yet many models used to study measurement error assume linear and homoskedastic errors. For example, top-coding of income causes nonlinear method effects \citep{gottschalk2010earnings}, and it is often thought that low earners over-report while high earners under-report, yielding ``mean-reverting'' random errors \citep[e.g.][]{kim2014response}. Third, the measurement quality of administrative data often differs over observations, yielding a mixture of measurement models. For example, the records may be obtained from a mixture of sources \citep{wallgren2007register}, such as both employer statements and employee self-reports, or the variable may be more ambiguously defined for some cases than for others: the income of day laborers is an example. Earlier approaches have not accounted for such heterogeneity. Currently, then, there is no generally applicable method to evaluate the extent of measurement error in register and survey data.


Our contributions to the literature are twofold: we present a framework for simultaneously estimating measurement error in register and survey data which addresses the shortcomings of earlier methods; we then provide guidance on the circumstances in which survey data or register data are preferable for use in research. Section \ref{sec:mtmm} introduces the modeling framework used to estimate the extent of measurement error in survey and register data simultaneously, and demonstrates how this framework encompasses existing methods. Section \ref{sec:application} applies the model to linked survey-register data on income and duration of employment from the German Federal Employment agency, while a simulation study in Section \ref{sec:simulation} evaluates the estimates obtained.

\section{Measurement error estimation from multiple error-prone sources}\label{sec:mtmm}

Our technique for simultaneously estimating measurement error in survey and administrative data builds on the ``multitrait-multimethod'' (MTMM) approach \citep{campbell_convergent_1959}. Given a set of variables of interest (``traits'') for which observed measurements exist in both the administrative data and a sample survey, our goal is to estimate the degree of measurement error in variables observed in both sources.

Let $y_{tm}$ denote an observed random variable measuring the $t$-th trait using the $m$-th method. In the application described here, $m$ will denote either administrative or the survey measurement.

\begin{example}\label{ex:mtmm-data}
	Suppose true income from full-time jobs $\eta_1$, part-time jobs $\eta_2$,
	and other types of jobs $\eta_3$ are of interest, for instance for future
	study of their relationship with educational attainment.
	Corresponding error-prone observed measures
	$y_{11}$, $y_{21}$, and $y_{31}$ are obtained in an administrative register. For a
	random subsample of cases, we also have survey measures of the same variables: $y_{12}$, $y_{22}$, and $y_{32}$. There are thus three ``traits'' (full-time, part-time, and
	other income) and two ``methods'' (register and survey), and six observed variables.
	An equivalent view is that $y_{tm}$ results from a repeated measures design in which
	the factors ``trait'' and ``method'' have been fully crossed.
\end{example}

\subsection{Current approaches to modeling MTMM data}
Commonly, MTMM data are analyzed using the linear model
\begin{equation}
	y_{tm} = \tau_{tm}  + 	\lambda_{tm} \eta_{t} +
		\gamma_{tm} \xi_{m} + \epsilon_{tm},
	\label{eq:mtmm-cfa}
\end{equation}
where $\tau_{tm}$ is the constant systematic bias in $y_{tm}$ and $\lambda_{tm}$ and $\gamma_{tm}$ are constant scaling factors with respect to the random variables. The ``trait factor'' $\eta_m$ is a random subject $\times$ trait interaction, symbolizing the ``true value'' of the trait measured by $y_{tm}$. The ``method factor'' $\xi_t$ is a random subject $\times$ method interaction, symbolizing method bias that differs over subjects but is common to variables measured with the same method. The residual $\epsilon_{tm}$ is random measurement error.

Assuming all $\eta_t$, $\xi_m$ and $\epsilon_{tm}$ follow a multivariate Gaussian distribution, Model \ref{eq:mtmm-cfa} is a confirmatory factor analysis (CFA) model with parameter vector $\param := (
\boldsymbol{\tau}',
\boldsymbol{\lambda}',
\boldsymbol{\gamma}',
\boldsymbol{\sigma}_{\boldsymbol{\eta}}',
\boldsymbol{\sigma}_{\boldsymbol{\xi}}',
\boldsymbol{\sigma}_{\boldsymbol{\epsilon}}')'$, where the parameters have been collected into vectors and $\boldsymbol{\sigma}_{\mathbf{x}}$ denotes the nonredundant elements of the covariance matrix of $\mathbf{x}$, stacked columnwise.


Under this model the implied product-moment correlation between two observed variables $y_{tm}$ and $y_{t'm'}$ (for $t \neq t'$) is
$$
	\text{cor}(y_{tm}, y_{t'm'}) =
	\begin{cases}
\lambda^*_{tm}\lambda^*_{t'm'} \text{cor}(\eta_t, \eta_{t'})  & \text{if}\; m\neq m'\\
		\lambda^*_{tm}\lambda^*_{t'm} \text{cor}(\eta_t, \eta_{t'}) +
	\gamma^*_{tm} \gamma^*_{t'm} &  \text{Otherwise},\\
		
	\end{cases}
$$
where $\lambda^*_{tm} = \lambda_{tm} [ \sigma_{\eta_t}(\param) / \sigma_{y_{tm}}(\param)]^{1/2} = \text{cor}(y_{tm}, \eta_{t})$ is the ``reliability coefficient'' of $y_{tm}$ and
$\gamma^*_{tm} = \gamma_{tm} [\sigma_{\xi_t}(\param) / \sigma_{y_{tm}}(\param)]^{1/2} = \text{cor}(y_{tm}, \xi_{m})$ is the ``method effect''.
Thus, when the measures have been obtained by \emph{different} methods, the correlation between two observed error-prone variables is attenuated by a factor $\lambda^*_{tm} \lambda^*_{t'm'} $ relative to the correlation between the ``true scores'' $\eta_{t}$ and $\eta_{t'}$: a classical result \citep[e.g.][]{lord_statistical_1968,fuller1987}.
This result shows that it is essential to model both random measurement error $\epsilon_{tm}$ and individual method biases $\xi_m$: their presence will have dramatically different effects on subsequent analyses of interest. The MTMM design allows for the separation of these two error factors.

This approach has led to a large literature on MTMM modeling using CFA (structural equation modeling) to estimate the degree of random and systematic measurement error in survey data \citep[e.g.][]{alwin1973approaches,andrews1984construct,saris_evaluation_1991,saris_design_2007,bakker2012estimating}. Extension for ordinal categorical data using the ``ordinal factor analysis'' model \citep{muthen1983categorical} have also been applied \citep{oberski2010categorization}. Recently, \citet{oberski2013latent} introduced a latent class factor \citep{vermunt_factor_2004} MTMM model.

\vspace{12pt}
The MTMM framework is in principle attractive for the modeling of measurement errors in administrative and survey data.
For register data, however, these currently available MTMM models are inadequate and can yield biased or nonsensical estimates, for the three reasons given in Section \ref{sec:intro}: nonnormality of true values, nonlinearity and heteroskedasticity of errors, and the existence of unknown groups that exhibit differential measurement error.
We generalize the MTMM framework to allow for these possibilities.

\subsection{The generalized multitrait-multimethod model}

We use generalized latent variable models \citep{skrondal2004generalized} to formulate a measurement model for MTMM data from an administrative register and a survey that can account for non-classical error processes, nonnormal distributions, and categorical data.
Generalized latent variable models are built up from (i.) linear GLM predictors; (ii.) GLM links and exponential family distributions; and (iii.) conditional independence relations. The conditional independence relations we use result from the MTMM design and are common to all MTMM models, whereas the choice of links and distributions is flexible: for this reason we call our approach a ``generalized multitrait-multimethod'' (GMTMM) model. The flexibility in links allows us to model nonlinearities and heteroskedasticities in the error process, while the choice of distributions for the latent variables allows for nonnormality of the true values.
Finally, when heterogeneous measurement error processes need to be accounted for, a finite mixture is used that allows the parameters of the linear predictors to differ over the mixture components.


\paragraph{(i.) Linear predictors.}

For continuous observed data, linear predictors for the observed variables $y_{tm}$ are:
\begin{equation}
	\nu_{tm} = \tau_{tm} + \lambda_{tm} \eta_{t} + \gamma_{tm} \xi_{m},
\end{equation}
where, for identification purposes, the first loading of each trait factor $\eta_t$ and method factor $\xi_m$ is often set to unity, $\lambda_{t1} = \gamma_{1m} = 1$. For categorical observed data, linear predictors for category $y_{tm} = k$ are
\begin{equation}
	\nu_{ktm} = \tau_{ktm} + \lambda_{ktm} \eta_{t} +  \gamma_{km} \xi_{m},
\end{equation}
where the first category can be chosen as a reference by setting $\tau_{1tm} = \lambda_{1tm}^{(\eta)} = \lambda_{1m}^{(\xi)} = 0$ \citep[e.g.][]{vermunt2013technical}.

The above linear predictors are common to all population units, and therefore assume that the measurement process is homogeneous. When the error process is thought to be heterogeneous, the linear predictor parameters are allowed differ over the mixture components, yielding an additional subscript $\nu_{tm,s}$ or (for categorical data) $\nu_{ktm,s}$.

\paragraph{(ii.) Links and distributions.}

Each of the observed and latent variables is assigned a  distributional ``family'' and a link function $g(\cdot)$ connecting the linear predictor to the expectation of the response $y_{tm}$ is chosen,
\begin{align}
	g(\text{E}[y_{tm} | \eta_t, \xi_m]) = 	\nu_{tm}, & & \text{or} & 	&
		g(\text{E}[y_{ktm} | \eta_t, \xi_m]) = 	\nu_{ktm},
\end{align}
depending on whether the observed variable is continuous or categorical.

We denote the choice of the conditional distribution of the observed responses given the latent variables as $f_y := p(y_{tm} | \eta_t, \xi_m)$ with parameter vector $\param_y$. Similary, the multivariate distribution of the latent ``true score'' variables is denoted $f_\eta$ with parameters $\boldsymbol{\theta}_\eta$ and the distribution of the latent ``method'' variables $f_\xi$ with parameters $\boldsymbol{\theta}_\xi$. Depending on whether the variables to which they refer are continuous or categorical, $f_y$, $f_\xi$ and $f_\eta$ may be probability density or probability mass functions.
Finally, the finite mixture components are assigned a multinomial distribution.

\paragraph{(iii.) Conditional independencies.}

The specification of the generalized latent variable model is completed with assumptions of conditional independence that are necessary for identification of the model parameters from observables. These assumptions mirror those of the linear MTMM model.

\begin{assumption}
	The observed variable $y_{tm}$ is conditionally independent of all other observed variables given its trait factor $\eta_t$ and method factor $\xi_m$.
	\label{assumption:conditional_independence}
\end{assumption}
Assumption \ref{assumption:conditional_independence} implies that the joint conditional distribution of observed given latent variables can be factored into the  univariate conditional distributions, i.e.
	\begin{equation}
		p(\mathbf{y} | \boldsymbol{\eta}, \boldsymbol{\xi}, \param) =
			\prod_{t,m} f_y(y_{tm} | \eta_t, \xi_m, \param_y).
	\end{equation}

\begin{assumption}
	The latent method factors $\boldsymbol{\xi}$ are mutually independent and independent of the trait variables $\boldsymbol{\eta}$.
	\label{assumption:method_independence}
\end{assumption}
Assumption \ref{assumption:method_independence} implies that the latent variable joint distribution can be factored into
\begin{equation}
p(\boldsymbol{\xi}, \boldsymbol{\eta} | \param) = f_\eta(\boldsymbol{\eta}| \param_\eta) \prod_m f_\xi(\xi_m | \param_\xi).
\end{equation}
Note that there may still be dependencies among the latent trait variables in the vector $\boldsymbol{\eta}$.

\begin{figure}\begin{center}
	\includegraphics[width=.8\textwidth]{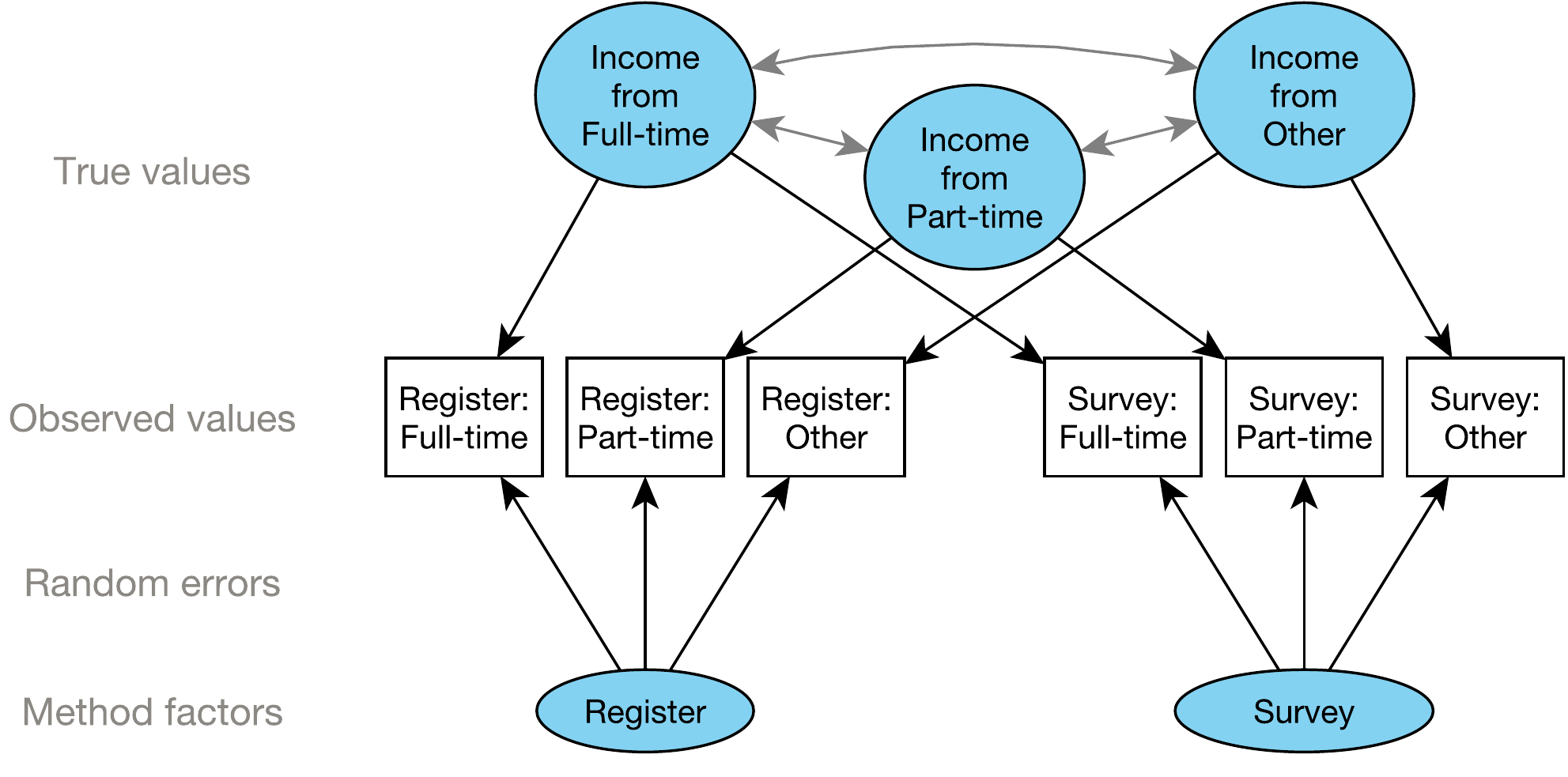}
	\caption{A generalized multitrait-multimethod (GMTMM) model for three ``traits'' using administrative data and a survey as measurement ``methods''. The example traits signify personal income from full-time, part-time, and other kinds of employment over a certain period.
		\label{fig:mtmm-register}
}
	\end{center}

\end{figure}

\paragraph{Example 1 (continued).}
The conditional independencies can be displayed in a graph with directed arrows for GLM regressions and undirected edges denoting possible (conditional) dependence.
Figure \ref{fig:mtmm-register} shows the GMTMM model for the six-variable MTMM data from Example \ref{ex:mtmm-data}.
In the Figure, observed variables $y_{tm}$ are shown as rectangles while unobserved random variables (factors) are shown as ellipses. Assumption \ref{assumption:conditional_independence} can be verified by noting that conditioning on the hidden nodes yields an independence graph \citep[e.g.][]{lauritzen1996graphical}.

\paragraph{Likelihood.}

When the error process is thought to be  homogeneous, the marginal likelihood $	p(\boldsymbol{y} | \param)$ is
\begin{equation}
	p(\boldsymbol{y} | \param) = \int\int{\left[
			 f_\eta(\boldsymbol{\eta} | \param_\eta)
		\prod_m	
			 f_\xi(\xi_{m} | \param_\xi)
		\prod_{t,m}
			f_y(y_{tm} |  \eta_{t}, \xi_{m}, \param_y)
	\right]} d \boldsymbol{\eta} d \boldsymbol{\xi}.
	\label{eq:model}
\end{equation}
where assumptions \ref{assumption:conditional_independence} and \ref{assumption:method_independence} are used and the integral is defined as a sum for discrete latent variable distributions.

For heterogeneous error processes, in which a mixture of error processes is thought to be present, define $p(\boldsymbol{y} | S, \param_{s})$ as the component-specific marginal likelihood, with component specific parameters $\param_{s}$. Typically, it is the measurement parameters that are thought to differ over components: that is,  the linear predictors $\nu_{tm,s}$.
We then introduce an unobserved discrete variable $S$ with categories equal to the number of components, so that the marginal likelihood of the observed data becomes
\begin{equation}
	p(\boldsymbol{y} | \param) =	\sum_{S} p(S) p(\boldsymbol{y} | S, \param_{s}).
		\label{eq:model-lca}
\end{equation}
Since the mixture proportions $p(S)$ are typically unknown, this implies an additional $|S|$ parameters in $\param$ to be estimated.

\subsection{Special cases of the GMTMM model}

By choosing different link functions, distributions, and error structures, a range of models that has been introduced in the literature to estimate measurement error in MTMM designs and administrative register data result as special cases of the GMTMM model.

\begin{example}
A common choice is to assume homogeneous errors, the identity link function $g(x) = x$, and distributions $f_y = \text{N}(\nu_{tm}, \sigma_{\epsilon_{tm}})$, with Gaussian latent variables $f_{\boldsymbol{\eta}} = \text{MVN}[\mathbf{0}, \boldsymbol{\Sigma}(\param_\eta)]$, $f_{\xi} = \text{N}(0, \sigma_{\xi_m})$, leaving $\boldsymbol{\Sigma}(\param_\eta)$  unrestricted so that $\param_\eta = \boldsymbol{\sigma}_{\boldsymbol{\eta}}$. This is the linear confirmatory factor analysis MTMM model presented above. This model was applied to linked survey-register data by \citet{bakker2012estimating} and \citet{scholtus2013estimating}.
\label{ex:CFA-MTMM}
\end{example}

\begin{example}\label{ex:ordinal-probit}
Leaving $f_\xi$ and $f_{\eta}$ unchanged from Example \ref{ex:CFA-MTMM},
the probit factor model for binary data results from choosing $f_y = \text{Binomial}[\text{E}(y_{tm})]$ with $g = \Phi^{-1}(\nu_{tm})$, where $\Phi$ is the standard normal distribution function. If, instead, the link function $g = \text{logit}(\nu_{tm})$ is chosen, a ``two-parameter logistic'' item response theory MTMM model is obtained.

Ordered categorical data can be modeled by choosing $f_y = \text{Multinomial}[\text{E}(y_{tm} = k)]$, redefining the observables, and choosing the link function $$g[\text{Pr}(y_{tm} \leq k | \eta_t, \xi_m)] = \Phi^{-1}(\nu_{ktm}),$$
where the loadings are set equal over categories, $\lambda_{ktm} = \lambda_{tm}$, $\gamma_{ktm} = \gamma_{tm}$, and the category-specific intercept $-\tau_{ktm}$ plays the role of a cumulative probit ``threshold'' \citep{rabe2004generalized}. An ordered probit
relationship between $y_{tm}$ and the latent variables is thus specified. This model is known as the ``ordinal factor analysis'' model in the structural equation modeling literature \citep{muthen1983categorical} and is a multidimensional version of the ``normal ogive graded response model'' in the item response theory literature \citep{samejima_estimation_1969}.
\label{ex:oCFA-MTMM}
\end{example}

\begin{example}
The CFA and categorical CFA models in Examples \ref{ex:CFA-MTMM} and \ref{ex:oCFA-MTMM} relied on normally distributed latent variables. It is possible to relax this assumption of normally distributed latent variables by specifying $f_{\eta} = \text{Multinomial}(\boldsymbol{\pi}_{\boldsymbol{\eta}})$ with free joint probability vector $\param_\eta = \boldsymbol{\pi}_{\boldsymbol{\eta}}$, and univariate distributions $f_{\xi} = \text{Multinomial}(\boldsymbol{\pi}_{\xi_m})$, with free univariate probability vectors $\param_\xi = \{\boldsymbol{\pi}_{\xi_m}\}$. The number of latent categories to which $f_\eta$ and $f_\xi$ refer must be chosen in advance, yielding a finite mixture or ``latent class'' MTMM model \citep{oberski2013latent}.
When accompanied by the choice $f_y = \text{Multinomial}$, this model was described as ``nonparametric'' by \citet[sec. 4.4.2]{skrondal2004generalized} and as ``semiparametric'' by \citet{heinen1996latent}.
\end{example}

\subsection{Estimation and identification of GMTMM model}\label{sec:estimation}



The parameters $\param$ can be estimated from linked survey-register data when there are at least three ``traits''--that is, variables of interest that have been measured with error in both survey and administrative register.
Standard estimation procedures for generalized latent variable models can be used to estimate the GMTMM model \citep[e.g.][chapter 6]{skrondal2004generalized}. The most general is to use standard optimization algorithms to maximize the marginal likelihood from Equation \ref{eq:model}  or \ref{eq:model-lca}. For certain models, such as latent class MTMM models, direct maximization of the marginal likelihood may become unstable. An expectation-maximization (EM) algorithm can then be used \citep{mclachlan2007algorithm}.

Many of the special cases of GMTMM models, including the examples given above, can be estimated using standard software for latent variable modeling such as Latent Gold \citep{vermunt2013technical} or GLAMM \citep{rabe2004generalized}, that implement this estimation strategy. Moreover, specialized efficient estimation procedures already exist for certain special cases of the GMTMM model. For example, the linear factor analysis MTMM model can be formulated as a covariance structure model with a closed-form marginal likelihood \citep{bollen_structural_1989}. The ordinal factor analysis (cumulative probit) model can be similarly dealt with by first computing polychoric correlation coefficients \citep{muthen1983categorical}. Such models can be fit using standard software for structural equation modeling. Other possible combinations of choices may require specialized software.

\subsection{Model identification}	

The GMTMM model is a latent variable model, and its parameters are therefore not necessarily identifiable. A first point of interest is whether a given GMTMM model, such as the ordinal CFA MTMM model (Example \ref{ex:ordinal-probit}), will have identifiable parameters. A second point of interest is what number of traits and methods are minimally required to identify the parameters of any GMTMM model. Assessing identifiability can be particularly relevant in advance of designing a survey to evaluate administrative data quality, since this will determine how many questions should be asked in the survey.

We take parameters to be ``identifiable'' if and only if a finite number of parameter values will lead to any given likelihood for all parameter values of nonzero measure \citep[see][for some of the subtleties involved in this definition]{allman2009identifiability}. Trivially, for example, with only one variable observed on one trait using a single method, it is clearly not possible to establish the parameter values regarding the latent trait and latent method factor variables separately, since infinitely many choices of $\param$ will lead to the same likelihood. On the other hand, the well-known ``label switching'' phenomenon in latent class-type models \citep{mclachlan2000finite} leads to finitely many solutions and is therefore not considered an identification problem here. Similarly, choices of $\param$ that lead to rank deficiencies but have a point mass in the parameter space \citep[see for example][]{shapiro1983investigation} are not considered identification problems in this definition.

First, under the definition given, a given GMTMM model's parameters will be identifiable if and only if the Jacobian $\partial p(\boldsymbol{y} | \param) / \partial \param$ is of full column rank almost everywhere \citep[Theorem 1]{catchpole1997detecting}. Equivalently, the rank of the information matrix may be examined. For GMTMM models with a closed-form marginal likelihood, this condition can be established analytically by assessing this rank using a symbolic algebra program. This may be considered an inconvenience by many applied researchers, however. For models without a closed-form marginal likelihood, analytical proofs are even more difficult. Numerical methods are then the more convenient tool to assess identifiability.

A common numerical approach is to examine the rank of the information matrix at the maximum likelihood estimate for a given dataset using the same software  used to fit the model. The disadvantage of this method is that it conditions on the data at hand. For example, a model may appear identified when it is not, due to boundary solutions, and it may appear non-identified for particular parameter values when it is identified in the larger parameter space.
To overcome this disadvantage, \citet{forcina2008identifiability} suggested evaluating the rank of the Jacobian at a large number of random values in the parameter space. This method has been implemented in the software Latent Gold 5 \citep{vermunt2013technical}.

\bigskip

This Section introduced a generalized multitrait-multimethod model that can be used to estimate measurement error when at least two separate measures of at least three different phenomena are available. The GMTMM model can deal with nonnormality of true values, nonlinearity and heteroskedasticity of errors, and the existence of unknown groups that exhibit differential measurement error.  It is therefore applicable to estimating measurement error in administrative register data and surveys simultaneously. It is also more generally applicable to situations where such error structures are thought to exist in multiple error-prone sources.

\section{Application to administrative data on income and duration of employment}\label{sec:application}





This Section applies the GMTMM model to a unique dataset provided by the Institute for Employment Research (\emph{Institut f\"ur Arbeitsmarkt- und Berufsforschung}, IAB), the research institute of the German Federal Employment Agency (\emph{Bundesagentur f\"ur Arbeit}, BA). The BA's normal operations include job placement and payment of benefits, and for these purposes it maintains an extensive database of citizens' (un)employment histories dating back to 1975. This database covers German employees who are subject to social security contributions as well as recipients of entitlements, comprising about 86\% of the overall German labor force. Excluded from the register are most civil servants, the self-employed, and others who have never been in contact with the Agency, such as the never-employed.

Both survey data and the BA's register data are routinely used for labor market and policy research--especially those on income and duration of employment.
For consenting respondents, we gained IRB approval to link administrative record data from the Agency with a telephone survey conducted by the IAB (IAB Besch\"aftigtenhistorik (BEH) Version 09.01.00, N\"urnberg 2012).
Restricted access to the anonymized linked survey-administrative data was provided at the Agency's offices; the raw data cannot be made publicly available for legal reasons.


Particularly of interest are the BA's records on \emph{income} from full-time, part-time, and  ``marginal'' employment. ``Marginal'' employment, also known as ``Minijobs'', is a common form of low-income employment in Germany, yielding monthly income of up to 400 Euro (at the time of data collection); at or below this maximum, the employee is exempt from income taxes and social security.
Of additional policy interest are the \emph{durations} of the last employment spell of these three employment types. These data are not provided by the employees themselves, but rather by their employers, who are legally required to report their employees' income accurately for the purposes of taxes, benefits, and social security.

However, exactly because the income and duration data were collected for the BA's administrative purposes, measurement error can become a serious issue for research in spite of reporting accuracy,  because measurement errors in administrative data need not come from the reporting itself \citep{bakker2009trek,groen2012sources}. For example, although the employers will presumably fulfill their mandate to report accurately, when compiling historical records there may be mismatches and time lapses in an individual's record. Similarly, smaller jobs may simply be absent from the records, again leading to a mismatch in ``last part-time job'', for instance.
These issues will lead to random and correlated measurement error for research purposes.

To obtain the survey measurement, a stratified sample of 2,400 respondents was asked to provide  information on income and employment duration from full-time, part-time, and marginal employment \citep[see][for further description of the sample design]{eckman2014assessing}.
 The survey had a response rate (AAPOR RR1) of 19.4\%. In the following analyses, we accounted for the sample stratification using complex sampling adjustments. Of the respondents, 2,284 (95\%) provided informed consent to record linkage between the survey and the administrative registers. This linkage  could be performed using unique person identifiers, so that it seems reasonable to assume no linkage errors were present. By linking the administrative data to the survey data, we thus obtained MTMM designs with three traits and two methods, one for each of the income and duration data.

The register provides income data only at the level of employment spells. This typically corresponds to an annual basis if a respondent was employed at the same employer throughout a given year. The survey, however, explicitly asks for the last monthly income from gainful employment which is the standard reference period used in most German surveys.
Assuming that salaries are paid evenly throughout the employment spell, the administrative data were converted to a monthly basis.

\subsection{Estimates of reliability and method effects in survey and administrative measures}

To demonstrate the flexibility of the GMTMM approach and account for possibly differing measurement processes in the two measures investigated, we fit different types of GMTMM models to the  duration and income data.

\paragraph{Duration data.}

For the duration data, we estimate Gaussian GMTMM models: that is, the familiar linear structural equation model using the standard SEM software \texttt{lavaan} for \texttt{R}  \citep{rosseel2013lavaan,Rlanguage}. The program code to estimate this model can be found in the Appendix.

\begin{figure}[tb]
	\includegraphics[width=\textwidth]{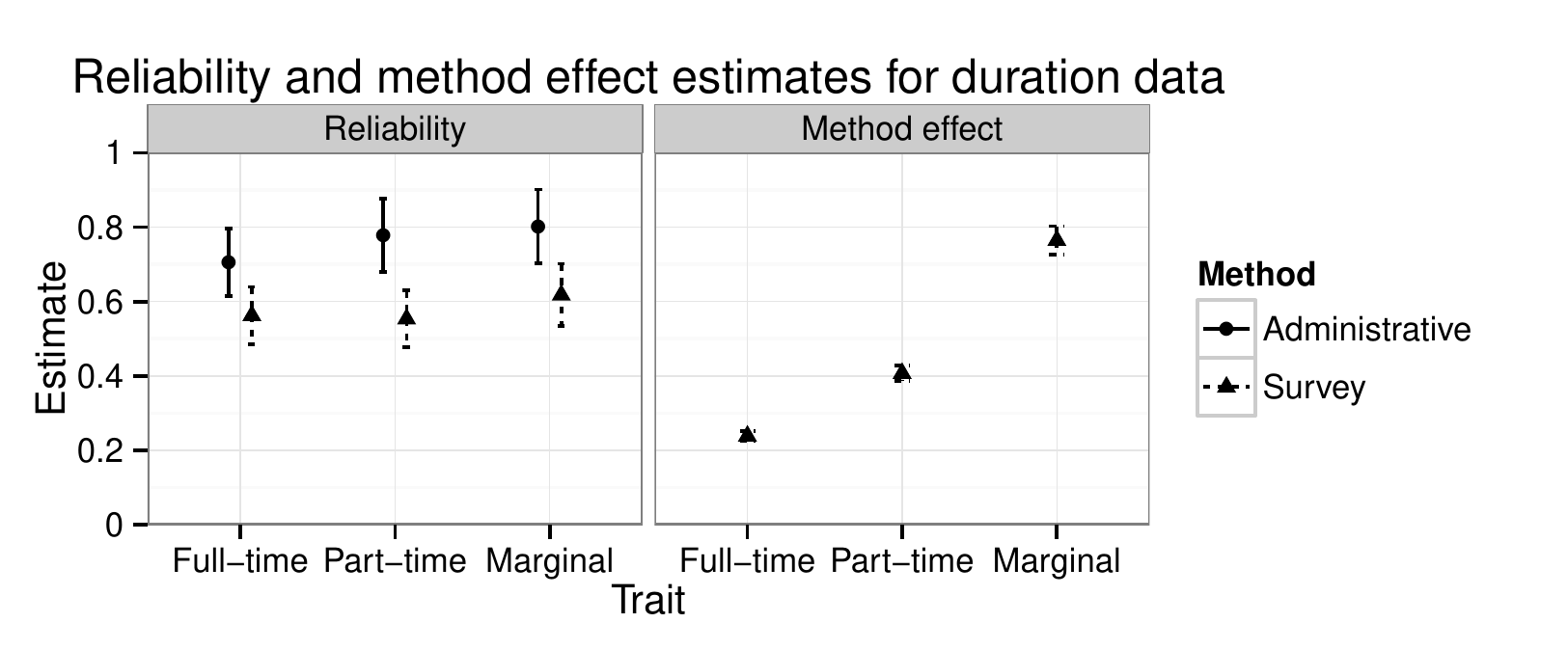}
	\caption{Reliability and method effect estimates for survey data, and reliability estimates for administrative register data on duration  of full-time, part-time, and ``marginal'' employment.}
	\label{fig:estimate-duration}
\end{figure}

This approach yielded estimates for the trait loadings ($\lambda_{tm}$), method loadings ($\gamma_{tm}$), factor (co)variances ($\sigma_{\xi m}$, $\sigma_{\eta t}$), and error variances ($\sigma_{tm}$). In a linear model, the quality of each administrative variable can be simply represented by two numbers: the reliability and the method effect. These represent, respectively, the correlation between the observed administrative variable and its measured trait, and between the observed variable and the method factor \citep{saris_design_2007}. A high reliability indicates that a survey question or register value contains little random error and accurately reflects the true value it measures. A high method effect, on the other hand, indicates that a substantial part of the variance is due to factors shared with other survey or register measures, but which are independent of the true values. An ideal measure would therefore have reliability one and zero method effect.
Estimates of the reliability and method effects are displayed for the duration data in Figure \ref{fig:estimate-duration}.



Figure \ref{fig:estimate-duration} shows reliability estimates in the left-hand panel and method effect estimates in the right-hand panel for the administrative and survey data on duration. The reliability estimates in Figure \ref{fig:estimate-duration} are between 0.7 and 0.8 for the administrative data, which indicates that reliability of the administrative data is acceptable, but far from perfect. For example, the correlation between administrative records on full-time duration and the person's true full-time duration is estimated at 0.7.  The administrative measures' reliabilities are clearly higher than the survey measures' reliabilities, which are around 0.6. Thus, the self-reports were somewhat less reliable than the administrative records, but neither measure was perfect.

 While fitting the model, the method effects ($\gamma_{tm}$) and method factor variances ($\sigma_{\xi m}$) for the administrative measures were estimated at zero but caused serious dependencies among the parameter estimates. We  followed \citet{eid_multitrait_2000} and \citet{saris_design_2007} in fixing these to zero and re-estimating the model without method dependencies in the administrative data. The right-hand panel of Figure \ref{fig:estimate-duration} therefore shows method effects for the survey measures only. These method effects can be seen as small for full-time durations, medium for the part-time durations, and very large for durations of ``marginal'' jobs. For example, a standardized method effect of 0.4 implies that answers to two survey questions on income will correlate by 0.4 above and beyond any true correlation between the two measures, thereby inflating relationship estimates that do not account for method effects. These large dependencies may be related to survey respondents' different but systematic interpretations of a ``duration'', or of what counts as a ``marginal'' job. However, there does not appear to to be any such effect in the administrative data.

\paragraph{Income data.}
To estimate the quality of the administrative register as well as the survey answers on income data, we adapt the model to recognize several aspects of the measurement process:
\begin{itemize}
	\item Following the econometrics literature \citep{tobin1958estimation}, censoring in income is accounted for;
	\item The relationship between true income and reported income is thought to be nonlinear \citep{kim2014response};
	\item Previous studies linking survey and register data \citep{scholtus2015validiteit} suggested that there is a subgroup of respondents for whom the two measures correspond exactly, whereas for others they do not, possibly suggesting a heterogeneous error process;
	\item There is a strong incentive to misreport one's income from a ``Minijob'' as being equal to or below 400 euros, since at the time of the survey this was the legal maximum income to qualify for tax exemption and social security exemption (see \S8 SGB [Social Security Code]).
\end{itemize}
Due to these factors, a linear Gaussian MTMM will not suffice. Instead, we choose $f_y$ to be the standard censored regression equation, use the ``nonparametric'' latent class factor analysis formulation of $f_\xi$ and  $f_\eta$ to allow for nonlinearity \citep{oberski2013latent}, and investigate whether an additional mixture component of $S$ in which the response is unrelated to the true value fits the data more closely than a homogeneous error structure. This model is no longer a standard structural equation model but can be estimated in the software for latent class (factor) analysis Latent GOLD 5.0 \citep{vermunt2013technical}. Program input   can be found in the Appendix.

The latent class factor analysis model does not impose a distribution on the latent trait and method factors, but instead approximates these distributions by  discrete interval-level latent variables whose category sizes are estimated from the data \citep{vermunt_factor_2004}. Moreover, the possibility of a heterogeneous error structure suggests the presence of an additional discrete nominal latent variable $S$. Since the number of categories for the latent trait, method, and error structure variables is unknown, we compare the fit of models with differing numbers of categories for each of these. Since increasing the number of categories of the method factors and the error structure variables beyond two never improved the model, we only show these comparisons for models with differing numbers of categories $K$ for the latent trait variables ($\eta_t$), with ($|S|=2$) and without ($|S|=1$) a heterogenous error structure.

 Table \ref{tab:model-fit} shows the fit of these models in terms of loglikelihood (LL), BIC, and AIC, as well as the number of parameters these models have. The model with three latent categories and a heterogeneous error process fit the data best in terms of BIC and AIC. This result suggests that there may indeed be differing error processes for different respondents. Since the model fit did not improve when increasing the number of latent categories from three to four, we selected the three-class heterogeneous model. In other words, we approximate the distribution of true latent income with a discrete three-category latent variable for which the category sizes are estimated. We also allowed for some proportion of the observations to be unrelated to the true value, for example because some fixed value (such as 400 euros) was always chosen in this group regardless of the true income.

\begin{table}
\begin{tabular}{rrrrrrrrrr}
\toprule
 & \multicolumn{8}{c}{Error process} \\

 & \multicolumn{4}{c}{Heterogeneous} &&   \multicolumn{4}{c}{Homogeneous} \\
 \cline{2-5}  \cline{7-10}
$K$ & LL & BIC & AIC & \# par. && LL & BIC & AIC & \# par. \\
\midrule
2 & -5060.0 & 10413.8 & 10195.9 & 38 && -5388.3 & 11024.0 & 10840.6 & 32 \\
3 & -4758.3 & \textbf{9825.9} & \textbf{9596.6} & 40 && -5272.1 & 10814.8 & 10614.1 & 35 \\
4 & -4848.9 & 10030.3 & 9783.8 & 43 && -5210.1 & 10714.1 & 10496.3 & 38 \\

\bottomrule
\end{tabular}
\caption{Fit of GMTMM models for the measurement error in administrative and survey data on income.}
\label{tab:model-fit}
\end{table}

\begin{table}\begin{small}

\begin{tabular}{llrrrrrrrr}

\toprule
 &&   \multicolumn{3}{c}{Trait} && \multicolumn{2}{c}{Method} && Overall \\
 \cline{3-5}  \cline{7-8}
 &  & 1 & 2 & 3 &  & 1 & 2 \\
 \midrule
	\multicolumn{6}{l}{\emph{Administrative	data	(log-income)}}\\
&Full-time	 & 1.11 & 2.69 & 4.31 &  &  &  &  & 1.85 \\
&Part-time	 & 0.65 & 1.54 & 2.45 &  &  &  &  & 1.08 \\
&Marginal	 & 0.09 & 0.23 & 0.36 &  &  &  &  & 0.21 \\
	\multicolumn{6}{l}{\emph{Survey	data	(log-income)}}\\
&Full-time	 & 2.20 & 3.16 & 4.12 &  & 5.52 & 2.25 &  & 2.65 \\
&Part-time	 & 0.91 & 1.67 & 2.45 &  & 1.44 & 1.26 &  & 1.28 \\
&Marginal	 & 0.27 & 0.33 & 0.38 &  & 0.33 & 0.32 &  & 0.32 \\
\bottomrule
\end{tabular}
\end{small}
\caption{Estimated relationships between categories of the latent trait variables $\eta$ and the expected observation of log-income from full-time, part-time, and marginal employment using the administrative and survey measures.}
\label{tab:income}
\end{table}

Table \ref{tab:income} shows the expected means of the administrative and survey measures of log-income for different categories of the latent trait and method variables. The table illustrates how the observed measures are estimated by the model to relate to the respective latent variables. The relationships in Table \ref{tab:income} are marginalized over the two categories of the error process latent variables $S$. Thus, the table shows how the relationship holds for a respondent whose error process is not known in advance. About 5\% (not shown in the table) are estimated to belong to the latent category in which a random value is given -- that is, a value that is unrelated to the trait or method variables.

\begin{figure}[tb]
	\includegraphics[width=\textwidth]{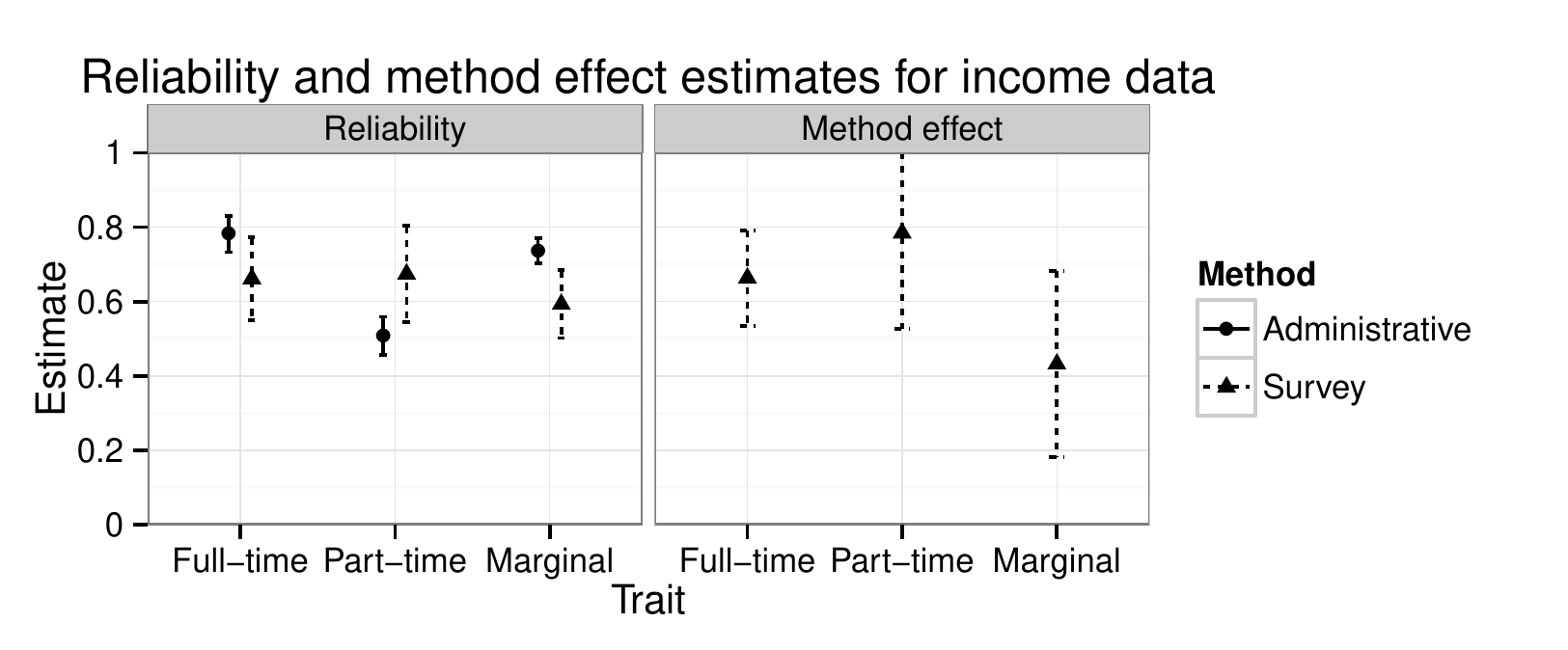}
	\caption{Reliability and method effect estimates for survey data, and reliability estimates for administrative register data on income from full-time, part-time, and ``marginal'' employment.}
	\label{fig:estimate-income}
\end{figure}

The model is no longer linear, so that reliability and method effect coefficients, which represent (linear) correlations are more difficult to interpret. However, it is possible to calculate the model-implied reliabilities $\text{cor}(y_{tm}, \eta_t)$ and method effects $\text{cor}(y_{tm}, \eta_m)$. These estimates, with confidence intervals based on bootstrapped standard errors, are shown in Figure \ref{fig:estimate-income}. The figure shows that while the administrative data on income from full-time and marginal jobs are estimated to be superior to the survey measures, the survey measure has a stronger linear correlation with true income level from part-time work. 
A possible explanation for this difference is a change in mandatory reporting procedures regarding part-time employment in the year 2011.
On the other hand, the survey measures do exhibit a strong method dependence, whereas again the administrative register measures were estimated to have no such method dependence.

\bigskip
In summary, we found for official administrative data obtained from the German Federal Employment Agency that the reliability of both survey \emph{and} administrative data was far from perfect. Estimated relationships between these observed variables and other variables of scientific interest will therefore be biased. Moreover, for some of these measures, method effects were found that will cause spurious dependencies where none exist among the true variables; when using administrative data, method dependence may be less of a concern. To prevent biases arising from measurement error in substantive analyses of income or duration data, correction methods for known error processes may be needed \citep[e.g.][]{saris_design_2007,vermunt2010latent,skrondal2012improved}.

\section{Simulation}\label{sec:simulation}

We demonstrate some key properties of the maximum likelihood estimates of GMTMM model parameter estimates using a simulation study. Since there are many possible GMTMM models that fall within this framework, we choose the model and parameter values based on the linked survey-register dataset obtained from the German Federal Employment Agency, and summarize bias and standard error accuracy under different conditions corresponding to sample sizes.

The response model chosen for the observed variables is a censored regression in which the unobserved trait and method variables are the regressors and the dependent variables are six observed indicators corresponding to the crossing of three traits and two methods. Thus, the response model for the observed variable $y_{tm}$ measuring trait $t$ with method $m$ is
\begin{equation}
	y_{tm} =   \begin{cases}
      0, & \text{if}\ y^*_{tm} \leq 0\\
      y^*_{tm}, & \text{otherwise}
    \end{cases},
\end{equation}
where $y^*_{tm}$ follows the linear factor model,
\begin{align}
	y^*_{tm} = \tau_{tm}  + \lambda_{tm} \eta_{t} +
		\gamma_{tm} \xi_{m} + \epsilon_{tm}, && \epsilon_{tm} \sim N(0, \sigma_{\epsilon, tm}).
\end{align}

The latent variables themselves are discrete interval-level variables with a multinomial distribution parameterized using the log-linear model
\begin{align}
	P(\eta_{1} = k_1, \eta_{2} = k_2, \eta_{3} = k_3) &=
		\frac{\exp\left(\mu_{k_1 k_2 k_3}
		\right)}{\sum_{k_1' k_2' k_3'}
		\exp\left(\mu_{k_1' k_2' k_3'}	\right)},\\
	P(\xi_{m} = k) &= \frac{\exp(\kappa_{mk})}{\sum_{k'} \exp(\kappa_{mk'})}
\end{align}
where
$\mu_{k_1 k_2 k_3} = \sum^{3}_{t=1} \alpha_{t k_t}+
				\phi_{12} \eta_{1,k_1} \eta_{2,k_2} + \phi_{13} \eta_{1,k_1} \eta_{3,k_3} + \phi_{23} \eta_{2,k_2} \eta_{3,k_3}$.

This model yields the following set of parameters, corresponding to the observed variable intercepts $\tau_{tm}$, trait loadings $\lambda_{tm}$, method loadings $\gamma_{tm}$, error variances $\sigma_{\epsilon, tm}$, as well as the latent variable loglinear intercepts $\alpha_{tk}$, and $\kappa_{tk}$  and latent loglinear associations $\phi_{tt'}$:
$$\param = \left(\{\alpha_{tm}\}, \{\kappa_{mk}\}, \{\tau_{tm}\}, \{\lambda_{tm}\}, \{\gamma_{tm}\}, \{\sigma_{\epsilon,tm}\}, \{\phi_{tt'}\}\right)^\prime$$
Furthermore, corresponding to the selected model from our application, we choose three categories for the latent trait and two for the latent method variables:
$$
	| \eta_t | = 3, 	| \xi_m | = 2.
$$

To ensure parameter values are realistic, we set them to the maximum-likelihood estimates found in our application, and vary the sample size across conditions, $n \in \{200, 500, 1000, 2000\}$. The results of simulating data from this model and analyzing them using the GMTMM model are summarized in Table \ref{tab:simulation}.

Table \ref{tab:simulation} summarizes the bias, defined as the difference between the true parameter value and the simulation average of the maximum likelihood estimate, as well as the ratio between and the ratio between the average simulation standard error and standard deviation over replications (``s.e./sd'').

It can be seen in Table \ref{tab:simulation} that under all conditions, the bias is small for most parameters and the estimated standard errors accurately reflect the simulation standard deviation. Exceptions to this good performance are the latent variable intercepts (e.g. $\alpha_{21}$ and $\kappa_{11}$) in the condition with the smallest sample size ($n = 200$). Although the bias in this condition is smaller for the other latent intercept parameters, there is a clear pattern of overestimating the size of the largest class and underestimating that of the other classes. This bias dissappears as the sample size grows larger. The other parameters do not appear to show any bias, even at the smallest sample size.

Table \ref{tab:simulation} also shows the performance of information-based standard errors as an estimate of simulation standard deviation. The standard errors perform well when sample size it at least 500. In the smallest sample size condition, some of the standard errors tend to underestimate the simulation standard deviation, which will lead to undercoverage of confidence intervals.

\bigskip

In summary, while the performance of the maximum-likelihood estimates is generally good, bias in some of the parameter estimates and many of the standard errors occurred when the sample size is small ($n=200$). Therefore, we recommend to use the GMTMM model with samples of at least 500 cases.

\begin{table}[tb]
\caption{Simulation results for a generalized MTMM model, under different sample sizes. Shown are the true values of the parameters, the simulation bias, and the ratio between the average simulation standard error and standard deviation over replications (``s.e./sd'').  }
\label{tab:simulation}
\centering\begin{footnotesize}
\begin{tabular}{rrlrrlrrlrrlrr}  \toprule
&&&\multicolumn{11}{c}{Sample size $n$}\\
\cline{4-14}
  & && \multicolumn{2}{c}{200} && \multicolumn{2}{c}{500} && \multicolumn{2}{c}{1000} && \multicolumn{2}{c}{2000} \\
  \cline{4-5}   \cline{7-8}   \cline{10-11}   \cline{13-14}
Par. & True &  & Bias & s.e./sd &  & Bias & s.e./sd &  & Bias & s.e./sd &  & Bias & s.e./sd \\
  \midrule
  $\alpha_{11}$ & 0.889 &  & 0.013 & 0.956 &  & -0.001 & 1.002 &  & -0.002 & 0.968 &  & -0.002 & 1.013 \\
  $\alpha_{12}$ & 0.085 &  & -0.009 & 1.001 &  & 0.004 & 1.088 &  & 0.008 & 1.067 &  & 0.004 & 0.994 \\
  $\alpha_{21}$ & 1.426 &  & 0.074 & 0.875 &  & 0.027 & 0.964 &  & 0.015 & 0.962 &  & 0.013 & 0.965 \\
  $\alpha_{22}$ & -0.305 &  & -0.013 & 0.943 &  & -0.002 & 0.999 &  & -0.010 & 1.020 &  & -0.006 & 0.985 \\
  $\alpha_{31}$ & -0.121 &  & 0.017 & 0.996 &  & -0.003 & 1.040 &  & -0.007 & 0.960 &  & -0.002 & 0.955 \\
  $\alpha_{32}$ & -0.356 &  & -0.007 & 0.948 &  & 0.008 & 1.015 &  & 0.010 & 1.021 &  & 0.006 & 1.069 \\
  $\kappa_{11}$ & 0.058 &  & 0.018 & 0.752 &  & 0.005 & 0.902 &  & 0.005 & 0.920 &  & 0.001 & 0.939 \\
  $\kappa_{21}$ & -0.888 &  & -0.015 & 0.917 &  & -0.008 & 0.967 &  & -0.003 & 0.940 &  & -0.005 & 1.001 \\
  $\tau_{11}$ & 1.296 &  & 0.001 & 0.940 &  & 0.003 & 0.963 &  & -0.000 & 1.042 &  & -0.001 & 1.013 \\
  $\lambda_{11}$ & 3.772 &  & -0.017 & 0.815 &  & -0.004 & 0.917 &  & -0.000 & 0.948 &  & 0.007 & 0.943 \\
  $\gamma_{11}$ & -1.025 &  & -0.007 & 1.047 &  & -0.003 & 1.022 &  & -0.004 & 1.105 &  & -0.002 & 0.983 \\
  $\tau_{21}$ & 0.693 &  & -0.015 & 0.943 &  & -0.000 & 1.049 &  & 0.004 & 1.065 &  & 0.003 & 1.096 \\
  $\lambda_{21}$ & 1.546 &  & 0.013 & 0.956 &  & -0.001 & 1.005 &  & -0.005 & 1.010 &  & 0.002 & 0.998 \\
  $\gamma_{11}$ & 0.043 &  & 0.031 & 0.850 &  & 0.008 & 0.953 &  & -0.000 & 0.973 &  & -0.003 & 0.954 \\
  $\tau_{31}$ & 0.366 &  & 0.001 & 0.870 &  & 0.000 & 0.988 &  & -0.000 & 0.943 &  & -0.000 & 0.991 \\
  $\lambda_{31}$ & -0.283 &  & -0.001 & 0.931 &  & -0.000 & 1.090 &  & 0.000 & 1.032 &  & 0.000 & 1.008 \\
  $\gamma_{31}$ & 0.001 &  & -0.001 & 0.830 &  & -0.001 & 0.961 &  & -0.000 & 1.050 &  & -0.000 & 1.061 \\
  $\tau_{12}$ & 4.811 &  & 0.004 & 1.025 &  & 0.000 & 1.015 &  & 0.005 & 1.014 &  & 0.004 & 0.950 \\
  $\lambda_{12}$ & 2.029 &  & 0.003 & 0.929 &  & -0.001 & 0.988 &  & -0.004 & 0.992 &  & -0.003 & 0.987 \\
  $\gamma_{12}$ & -3.169 &  & -0.003 & 1.026 &  & 0.002 & 1.023 &  & -0.001 & 1.038 &  & -0.002 & 0.958 \\
  $\tau_{22}$ & 1.017 &  & 0.009 & 0.915 &  & 0.002 & 0.982 &  & -0.001 & 0.947 &  & 0.002 & 0.968 \\
  $\lambda_{22}$ & 1.964 &  & -0.003 & 0.981 &  & -0.001 & 1.020 &  & 0.001 & 0.960 &  & 0.002 & 0.970 \\
  $\gamma_{22}$ & -0.224 &  & -0.002 & 0.902 &  & 0.001 & 1.019 &  & 0.003 & 0.966 &  & -0.000 & 0.967 \\
  $\tau_{32}$ & 0.384 &  & 0.001 & 0.959 &  & -0.000 & 0.945 &  & 0.000 & 0.968 &  & 0.001 & 1.094 \\
  $\lambda_{32}$ & -0.114 &  & -0.002 & 0.971 &  & -0.000 & 0.943 &  & -0.000 & 0.961 &  & -0.001 & 0.998 \\
  $\gamma_{32}$ & -0.006 &  & -0.001 & 0.963 &  & -0.001 & 0.995 &  & -0.000 & 1.006 &  & -0.001 & 1.099 \\
  $\phi_{12}$ & 2.916 &  & 0.067 & 0.882 &  & 0.032 & 1.001 &  & 0.020 & 0.969 &  & 0.009 & 0.986 \\
  $\phi_{13}$ & -0.992 &  & -0.012 & 0.895 &  & -0.033 & 0.950 &  & -0.008 & 0.912 &  & -0.000 & 0.997 \\
  $\phi_{23}$ & -0.289 &  & 0.059 & 0.872 &  & 0.020 & 0.986 &  & 0.005 & 1.016 &  & 0.012 & 0.998 \\
    $\sigma_{\epsilon,11}$ & 0.175 &  & 0.004 & 0.771 &  & 0.001 & 0.934 &  & -0.001 & 1.005 &  & -0.001 & 0.984 \\
    $\sigma_{\epsilon,21}$ & 0.420 &  & -0.017 & 0.993 &  & -0.007 & 0.971 &  & -0.004 & 1.055 &  & -0.003 & 1.074 \\
    $\sigma_{\epsilon,31}$ & 0.003 &  & -0.000 & 0.891 &  & -0.000 & 1.031 &  & -0.000 & 0.932 &  & -0.000 & 0.941 \\
    $\sigma_{\epsilon,12}$ & 0.545 &  & -0.005 & 1.043 &  & -0.005 & 0.931 &  & -0.002 & 0.940 &  & -0.002 & 0.980 \\
    $\sigma_{\epsilon,22}$ & 0.141 &  & -0.002 & 1.067 &  & 0.001 & 1.043 &  & -0.000 & 1.064 &  & 0.000 & 0.954 \\
    $\sigma_{\epsilon,32}$ & 0.015 &  & -0.000 & 1.030 &  & -0.000 & 0.993 &  & -0.000 & 1.039 &  & -0.000 & 1.081 \\
   \bottomrule
\end{tabular}\end{footnotesize}
\end{table}

\section{Discussion and conclusion}\label{sec:conclusion}

We showed how the quality of survey and administrative data can be evaluated using generalized multitrait-multimethod (GMTMM) models. This approach is an improvement over existing methods, which assume that either the survey or the administrative data are perfect measures. A general framework for data quality evaluation was introduced. This framework is more suited than existing MTMM approaches to administrative data particularities such as categorical measurement, nonlinearities, heterogeneous error processes, and nonnormality. We demonstrated the use of GMTMM models by applying them to administrative and survey data on income and duration of employment from the German Federal Employment Agency. A simulation study demonstrated good properties of the maximum-likelihood estimates for a GMTMM model with moderate sample sizes.

A clear advantage of our approach is that it allows for the presence of measurement error in both the survey and the administrative register. Furthermore, using the administrative register as a second measure in the MTMM design has an additional advantage over classical MTMM designs using repeated survey measures. When repeated survey measures are used,
survey respondents must answer questions on the same topic twice and may remember their answer, creating dependencies that are not modeled \citep{alwin2011evaluating}, although \citet{vanMeurs1995memory} provided some evidence that this might not occur in practice when sufficient time is allowed between the repetitions. The problem of memory bias does not occur, however, when the measurement methods are administrative and survey data collected separately. Therefore, besides allowing for the estimation of measurement error in administrative records,  the MTMM design using linked survey-register data is an  attractive method of estimating measurement error in survey variables.

Some limitations of our work remain.
First, we did not discuss model fit evaluation. However, this issue is not specific to GMTMM modeling, so that the standard machinery available for global and local fit assessment in generalized latent variable models can trivially be applied to GMTMM modeling \citep[see, e.g.][]{skrondal2004generalized,Oberski:WP:model-based-GOF,oberski:WP:lca-bvr}.
Second, little is known about the small sample properties of GMTMM model estimates. While simulation results by \citet{scholtus2013estimating} on the linear MTMM model were positive, other types of GMTMM models were not evaluated as to their stability and robustness. This remains a topic for future research.
Finally, in our application on German data, unique identifiers were available that allowed for close linkage between the survey and register. In other applications, however, such identifiers may not be available for legal reasons or they may not exist. In such cases, linkage error will occur as well as measurement error. Incorporating such errors into the GMTMM model remains a topic for future study as well.

\bibliography{iab-mtmm}

\end{document}